\documentclass{elsart}

\usepackage{amssymb}
\usepackage{graphicx}

\def\BibTeX{{\rm B\kern-.05em{\sc i\kern-.025em b}\kern-.08em
    T\kern-.1667em\lower.7ex\hbox{E}\kern-.125emX}}

\begin{document}

\begin{frontmatter}

\title{How does noise affect the structure of a chaotic attractor: A recurrence network perspective}

\author[label1]{Rinku Jacob}
\ead{$rinku.jacob.vallanat@gmail.com$}
\author[label1]{K. P. Harikrishnan\corauthref{cor1}}
\ead{$kp_{\_}hk2002@yahoo.co.in$}
\author[label2]{R. Misra}
\ead{$rmisra@iucaa.ernet.in$}
\author[label3]{G. Ambika}
\ead{$g.ambika@iiserpune.ac.in$}

\corauth[cor1]{Corresponding author: Address: Department of Physics, The Cochin College,  Cochin-682002, India; Phone No.0484-22224954;  Fax No: 91-22224954.} 

\address[label1]{Department of Physics, The Cochin College, Cochin-682002, India}
\address[label2]{Inter University Centre for Astronomy and Astrophysics, Pune-411007, India}
\address[label3]{Indian Institute of Science Education and Research, Pune-411008, India}

\begin{abstract}
We undertake a preliminary numerical investigation to understand how the addition of white and colored noise 
to a time series affects the topology and structure of the underlying chaotic attractor. We use the 
methods and measures of recurrence networks generated from the time series for this analysis. We explicitly 
show that the addition of noise destroys the recurrence of trajectory points in the phase space. By 
using the results obtained from this analysis, we go on to analyse the light curves from a dominant 
black hole system and show that the recurrence network measures are effective in the analysis of 
real world data involving noise and are capable of identifying the nature of noise contamination 
in a time series.
\end{abstract}


\end{frontmatter}

\section{Introduction}
It is now well known that many time evolutions in Nature are inherently governed by nonlinear dynamical 
systems. For a proper understanding of these time evolutions, one often resorts to the methods and tools 
of nonlinear time series analysis \cite {kan}, since the information regarding the system in most cases 
is available in the form of a time series. There are two important properties that are the hallmarks of 
every dynamical system - determinism and recurrence. The former implies that the future behavior of the 
system can be accurately predicted, given sufficient knowledge of the current state of the system. By 
the latter property, the trajectory of a dynamical system tends to revisit every region of the phase 
space over an interval of time \cite {eck}. Hence, these two properties are very important in nonlinear 
analysis of time series data. Of special interest, in the analysis, is the search for deterministic chaos 
in the time evolution of the system and the presence of an underlying chaotic attractor.  
Because of this, many quantifiers from chaos theory \cite {hil,spr} are constantly being employed in 
the nonlinear analysis of observational data. 

The single largest problem in the time series analysis of real data is the presence of noise, both white 
and colored, that tend to destroy both the above mentioned properties  of any dynamical 
system underlying the time series. Several aspects of the effect of noise on synthetic as well as real 
world data and on the quantifying measures of discrimination have been addressed by many authors since the 
advent of chaos theory some four decades back. For example, the influence of white noise on the 
logistic attractor \cite {may}, the effect of colored noise on chaotic systems \cite {rad}, method to 
distinguish chaos from colored noise in an observed time series \cite {ken} and the seminal work on the 
correlation dimension analysis of colored noise data by Osborne and Provenzale \cite {osb}, to mention a 
few. The above studies resulted in the development of surrogate methods \cite {sch1,sch2} to discriminate 
chaos from random noise in real world data. More details on the effect of noise on chaotic systems and 
discriminating measures can be found in several standard books on chaos \cite {kan,hil,spr}.

In this paper, we apply the newly emerged tool of recurrence networks (RN) and the related measures 
\cite {mar1,don1} to undertake a preliminary numerical analysis to show how a chaotic attractor is 
affected by white and colored noise. The basic idea of RN analysis is that the information inherent in a 
chaotic time series is mapped onto the domain of a complex network using a suitable scheme. One then uses 
the statistical measures of the complex network to characterize the underlying chaotic attractor in the 
time series. There are two aspects of the RN that make them special for the analysis of time series data. 
Firstly, since the network measures can be derived from a small number of nodes in the network, the method 
is suitable for the analysis of short, non stationary data \cite {don2}. Secondly, the type of RN called 
the $\epsilon$ - RNs (whose details are given in the next section) that we consider in this work, generally 
preserve the topology of the embedded attractor from the time series \cite {don3}.  We show this specifically 
for the standard Lorenz attractor below. Hence, the topological changes in the underlying attractor due to 
noise addition are also reflected in the corresponding RN. This, in turn, implies that the RN and measures 
can be effectively used to study how the topological structure of a chaotic attractor is affected by 
increasing levels of noise contamination.

\begin{figure}
\begin{center}
\includegraphics*[width=16cm]{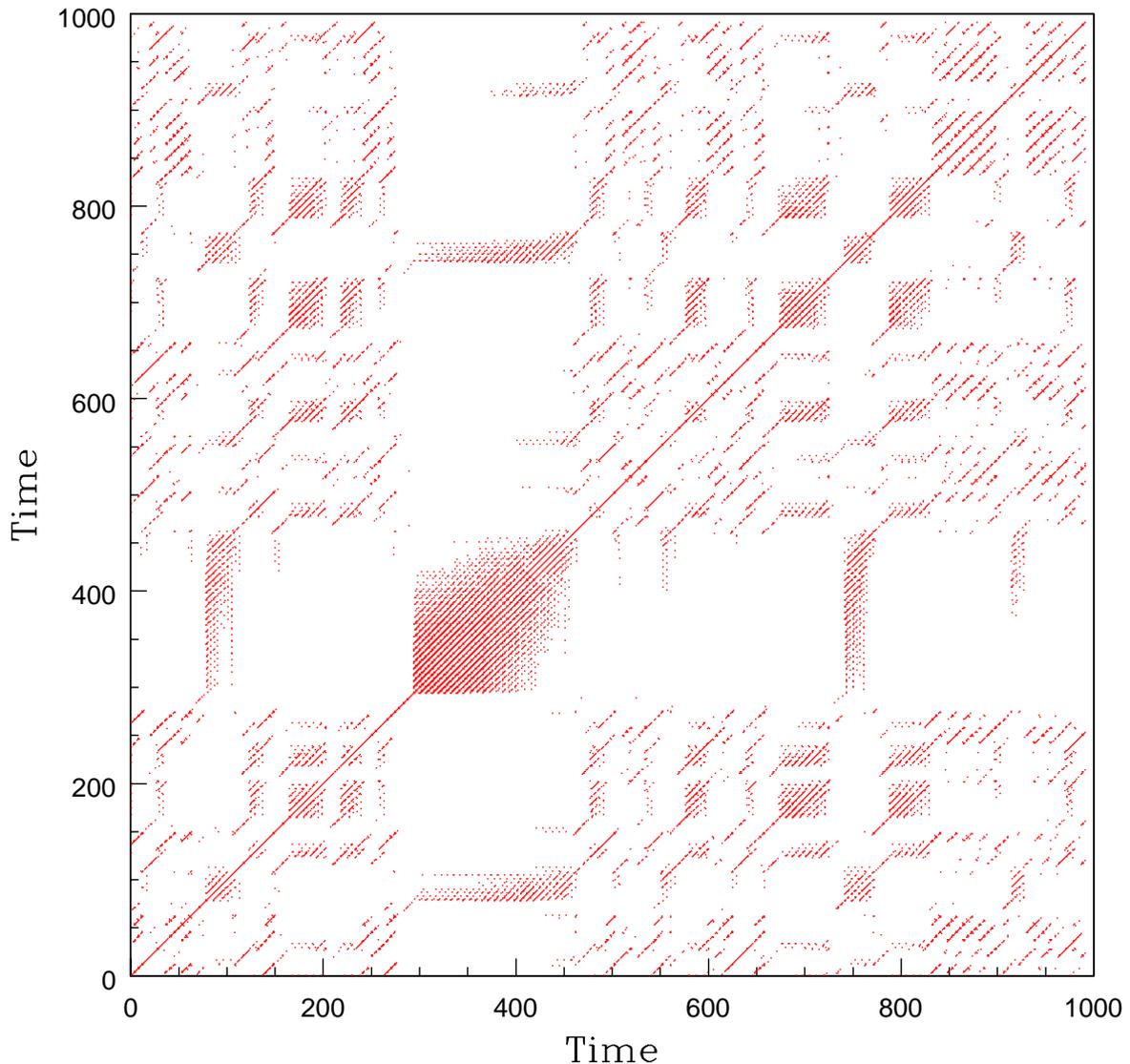}
\end{center}
\caption{The RP generated from the standard Lorenz attractor time series.}
\label{fig1}
\end{figure}

\begin{figure}
\begin{center}
\includegraphics*[width=16cm]{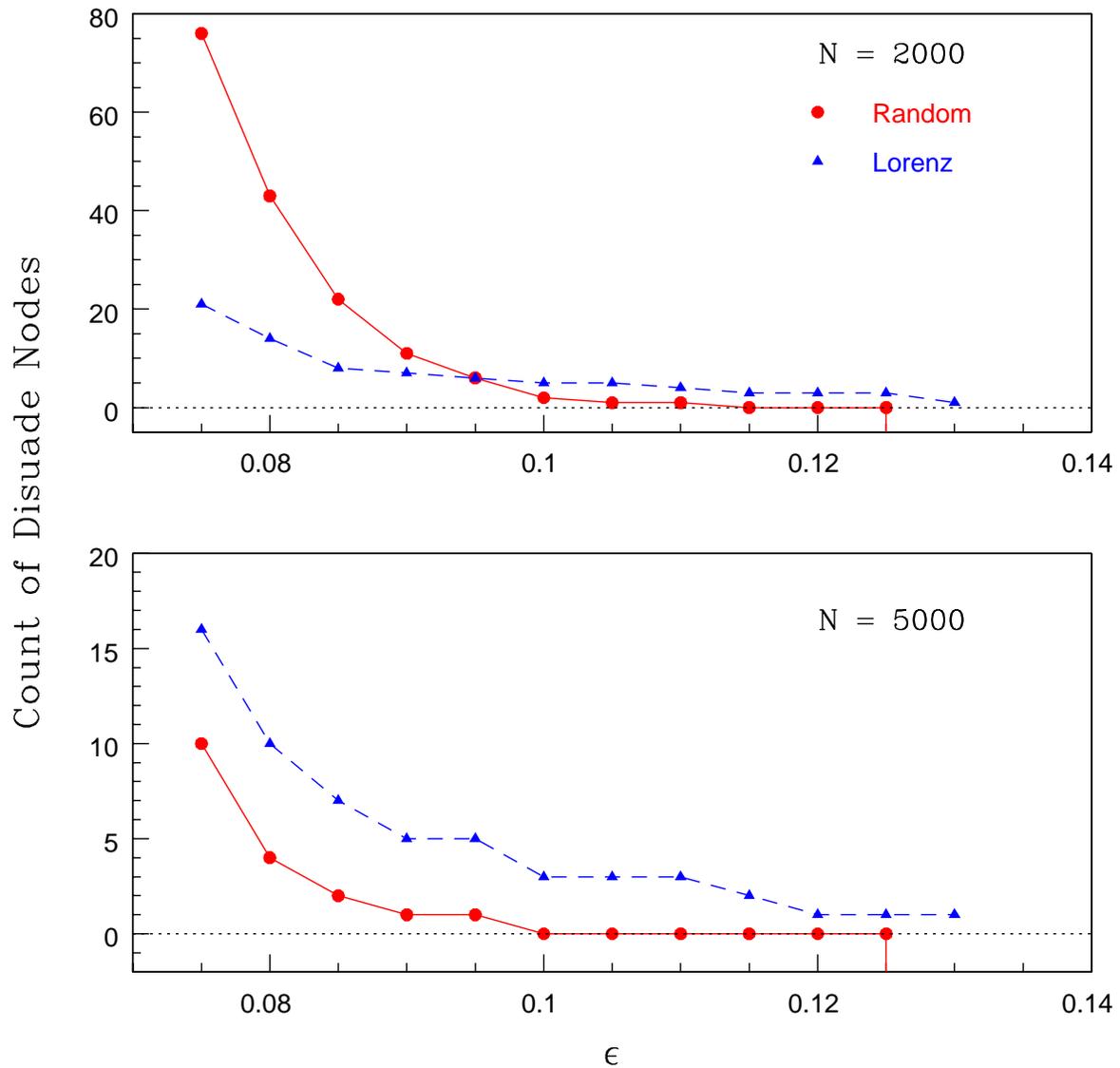}
\end{center}
\caption{Variation of the count of disuade nodes as a function of $\epsilon$ for the RNs constructed 
with $M = 3$ from random and Lorenz attractor time series for $N = 2000$ and $5000$.}
\label{fig2}
\end{figure}

The above properties of the RN have resulted in a number of practical appliations ranging from 
identification of dynamical transitions in model systems and real data \cite {dong,zou} to classification of 
cardio vascular time series \cite {avi}. However, there is one area where the RN measures have not been tested 
properly. For example, a systematic analysis of how the RN measures change with the addition of noise and 
how effective these measures are in the analysis of real data involving noise, are missing. To our knowledge, 
there is only one related study by Thiel et al. \cite {thi} using recurrence plot (RP) measures in this 
regard. These factors motivate us for the present analysis.

It should be noted that our aim in this analysis is neither to quantify the amount of noise in a given time 
series nor to distinguish deterministic nonlinear behavior from randomness in a given time series using any 
quantifying measure from RN. We focus on the following two aspects:

i) How the contamination of white and colored noise affect the topological structure of a low dimensional 
chaotic attractor?

ii) How effective are the RN measures in the analysis of real world data containing noise?

We introduce an additional measure for this purpose derived from the RN, which we call the \emph{k-spectrum.} 
Also, we use the time series from the standard Lorenz attractor as the prototype to study the effect of noise 
on synthetic data. 
The analysis is done by adding different amounts of white and colored noise to the Lorenz data. The results 
obtained from this is used for the analysis of real world data. 

Our paper is organised as follows: In the next section, we discuss the details regarding all the numerical 
tools used in this analysis, namely, the RP and the RN and the associated measures. In \S 3, we use the 
time series from the standard Lorenz attractor as an example to construct the RN and study how the network 
measures change by the addition of different amounts of white and colored noise. The \S 4 is devoted to the 
analysis of a few light curves from a standard black hole system GRS 1915+105 which are expected to contain 
different levels of white and colored noise. Conclusions are given in \S 5.

\begin{figure}
\begin{center}
\includegraphics*[width=16cm]{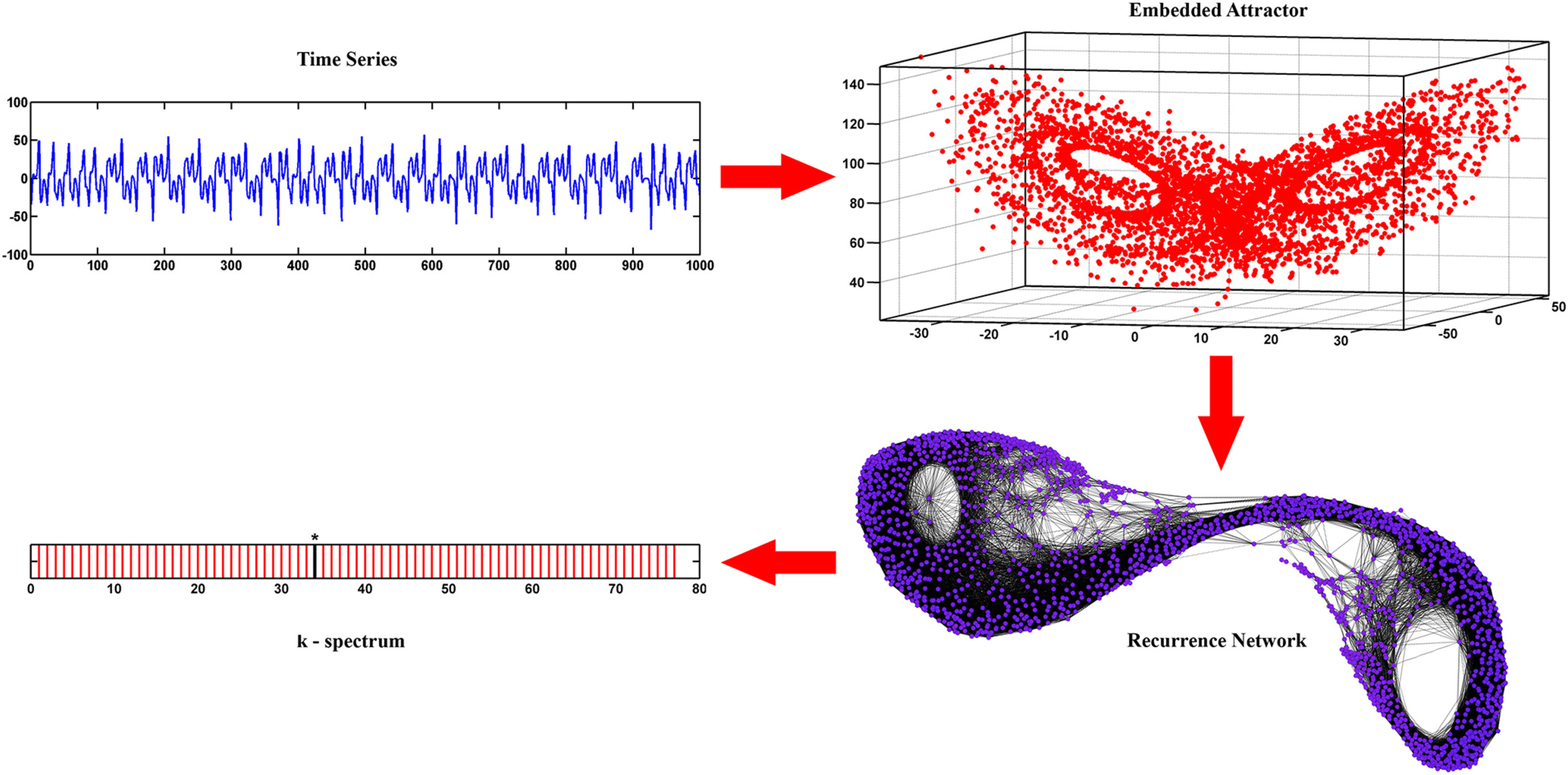}
\end{center}
\caption{The figure shows the construction of the RN from a time series. The top panel shows the 
Lorenz attractor time series and the embedded attractor from it using the time delay. The bottom 
panel shows the RN constructed from the embedded attractor and the corresponding k-spectrum 
(see text).}
\label{fig3}
\end{figure}

\begin{figure}
\begin{center}
\includegraphics*[width=16cm]{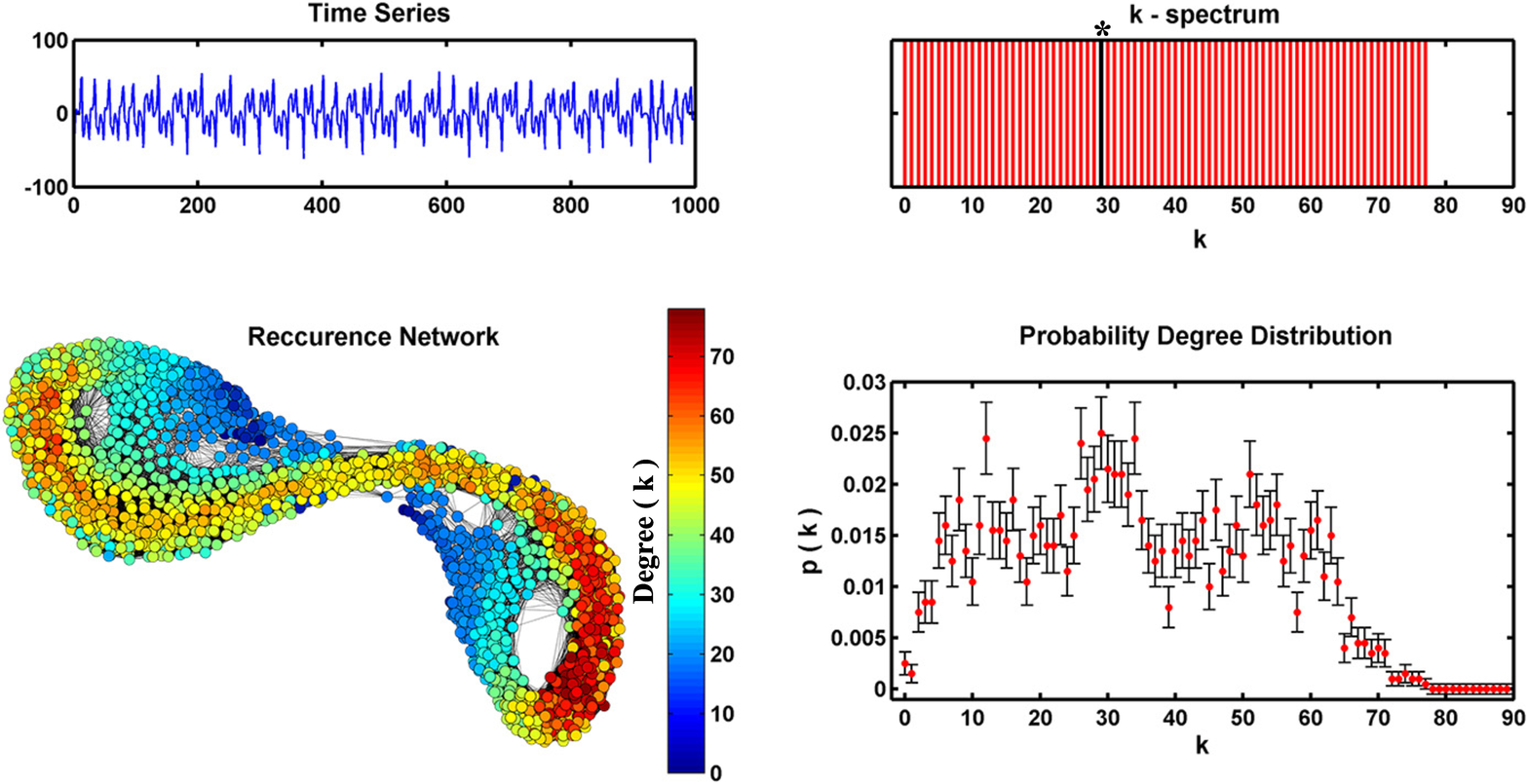}
\end{center}
\caption{The left panel shows the time series and the RN of the Lorenz attractor with number of 
nodes $N = 2000$. The right panel shows the k-spectrum and the degree distribution derived 
from the RN. Note that the variation in the degree of each node and its position in the RN 
can be clearly seen from the colour gradient for $k$.}
\label{fig4}
\end{figure}

\section{Numerical Tools: RP and RN}
In this section, we provide details of all the quantifying measures used for this study. RP is a visualisation 
tool introduced by Eckmann et al. \cite {eck}. It is a two dimensional graphical representation of the 
trajectory of the dynamical system in the form of a binary, symmetric matrix $\mathcal R$ where   
$R_{ij}$ = 1 if the state $\vec {x_j}$ is a neighbour of $\vec {x_i}$ in phase space and 
$R_{ij} = 0$, otherwise.  The neighbourhood is defined through a certain recurrence threshold $\epsilon$. 
To construct the RP from a scalar time series $s(1), s(2), .....s(N_T)$, the time series is first  
embedded in $M$-dimensional space using the time delay co-ordinates 
\cite {gra} using a suitable time delay $\tau$, where $N_T$ is the total number of points in the time 
series. The procedure creates delay vectors in the embedded space of dimension $M$ given by
\begin{equation}
\vec {x_i} = [s(i),s(i+\tau),...... s(i+(M-1)\tau)]
  \label{eq:1}
\end{equation} 
There are a total number of $N = N_T - (M-1)\tau$ vector points in the reconstructed space representing 
the attractor. Any point $\jmath$ on the attractor is considered to be in the neighbourhood of a reference 
point $\imath$ if their distance in the $M$-dimensional space is less than the threshold $\epsilon$. 
Thus we have 
\begin{equation}
R_{ij} = H (\epsilon - ||\vec {x_i} - \vec {x_j}||)
  \label{eq:2}
\end{equation}
where $H$ is the Heaviside function and $||..||$ is a suitable norm. Here we use the Euclidean norm. 
As an example, the RP constructed from the standard Lorenz attractor time series is shown in Fig.~\ref{fig1}. 

The RP has become more popular with the introduction of the recurrence quantification analysis (RQA) 
\cite {mar2,ssc} that uses the quantifying measures derived from the RP for the analysis of the data. For 
example, two important measures that are commonly used are the recurrence rate (RR) and the 
determinism (DET) given by
\begin{equation}
RR = {{1} \over {N^2}} \sum_{i,j} R_{ij}
  \label{eq:3}
\end{equation}

\begin{equation}
DET = {{\sum_{l \geq l_{min}}^{l_{max}} l p(l)} \over {\sum_{l=1}^{l_{max}} l p(l)}}
  \label{eq:4}
\end{equation}   
The former is a measure of the overall probability that a certain state recurs while the latter is a 
measure based on the distribution of the diagonal structures $p(l)$ and reflects how predictable the 
system is. Here we use DET as a quantifying measure in our analysis.

\begin{figure}
\begin{center}
\includegraphics*[width=16cm]{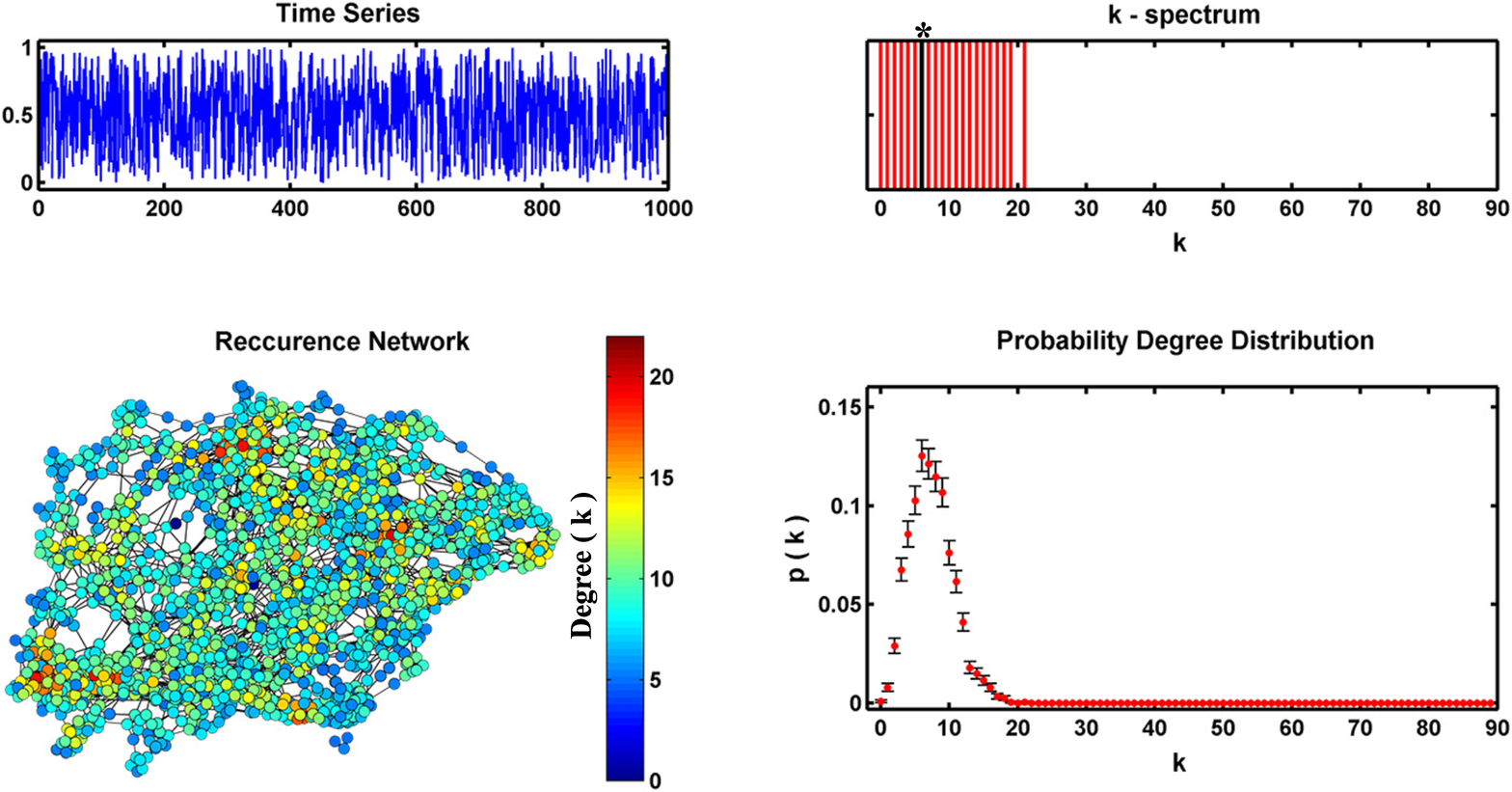}
\end{center}
\caption{The time series, RN, k-spectrum and the degree distribution for a pure random 
time series.}
\label{fig5}
\end{figure}

\begin{figure}
\begin{center}
\includegraphics*[width=16cm]{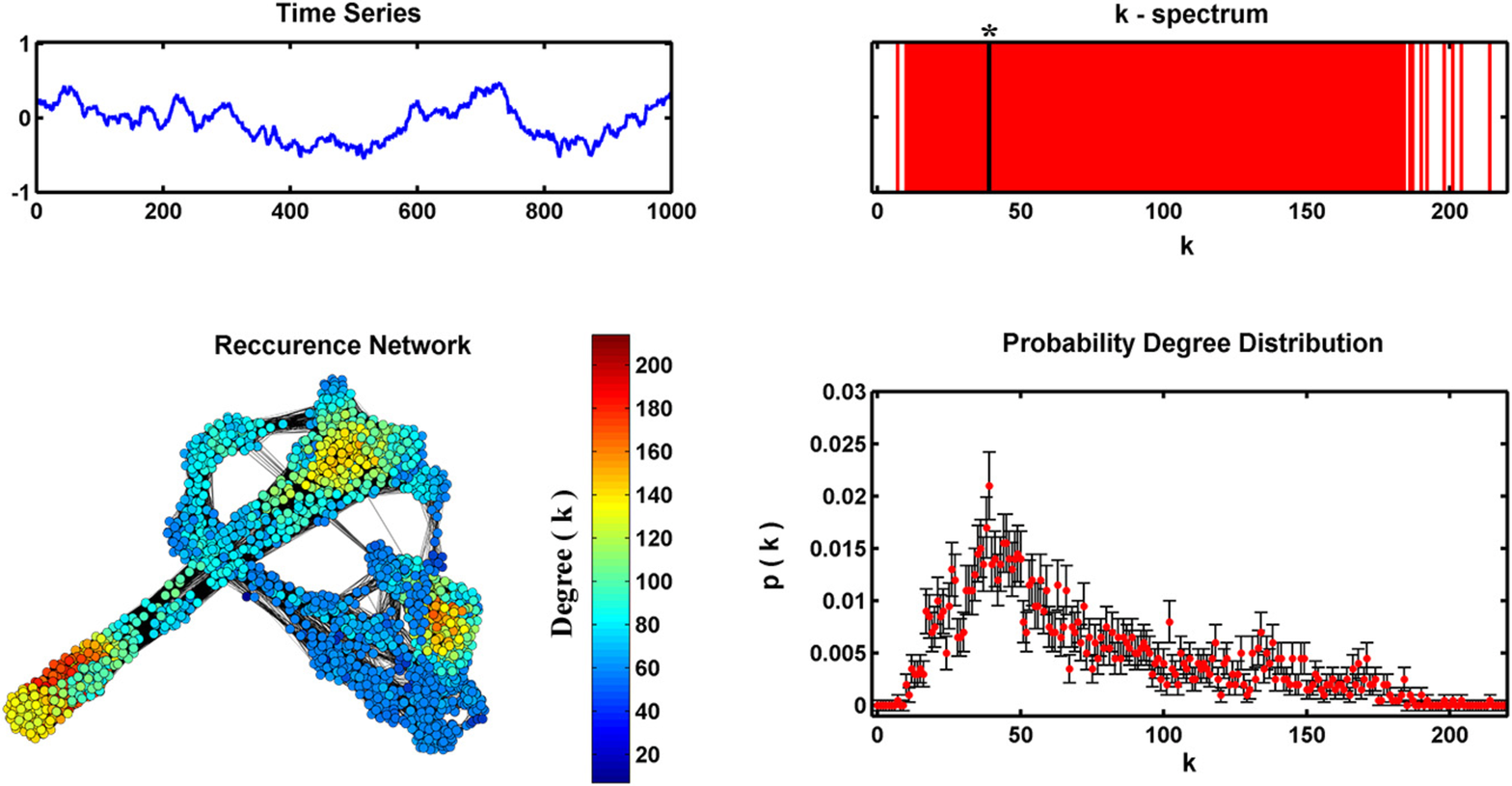}
\end{center}
\caption{Same as the previous figure for the red noise. Note that the range of the 
k-spectrum is much large compared to the RN of both random and Lorenz attractor and 
the topology of the RN is similar to that of a typical chaotic attractor.}
\label{fig6}
\end{figure}

The measures based on complex networks have emerged as a popular tool for the analysis of dynamical 
systems, especially in the form of time series, in the last two decades. The advantage of this approach 
is that one is able to extract information regarding the topological properties of the underlying 
attractor, which are otherwise unable to get, using conventional nonlinear time series analysis. 
The networks based on recurrence are called proximity networks which are mainly of two class, the 
fixed mass (or k-nearest neighbour) networks \cite {xu,sma} and the fixed volume 
(or $\epsilon$ - recurrence) networks \cite {don2,don3}. In the former, the neighburhood of a trajectory 
point for recurrence is defined in terms of fixed number of nearest neighbours while in the latter it is 
defined in terms of fixed phase space volume. In this paper, we use the procedure based on the 
$\epsilon$ - RN for the construction of networks from time series, confining to the case of unweighted 
$\epsilon$ - RN. 

The generation of $\epsilon$ - RN (which, from now on, we simply call RN) is closely associated with the RP. 
Infact, the adjacency matrix $\mathcal A$ for the unweighted RN can be 
obtained by removing the identity matrix from the recurrence matrix defined above:
\begin{equation}
A_{ij} = R_{ij} - \delta_{ij}
  \label{eq:5}
\end{equation}
where $\delta_{ij}$ is the Kronecker delta. Note that, once the adjacency matrix is defined, the time 
series has been converted into a complex network. Each point on the embedded attractor is taken as a 
node and it is connected to every other node whose distance is $\leq \epsilon$. The binary 
adjacency matrix contains 1 for all the pairs of points on the attractor that satisfy the condition 
$||\vec {x_i} - \vec {x_j}|| \leq \epsilon$ and zero otherwise.

The two important parameters associated with the RN construction from time series are the threshold 
$\epsilon$ and the embedding dimension $M$. In this paper, we do the analysis by choosing $M = 3$. 
For the selection of the critical threshold $\epsilon_c$ for RN construction, we closely follow the 
methods adopted by Donges et al. \cite {dong2} and Eroglu et al. \cite {erog}. The basic criterion employed 
is that the RN should appear as a giant cluster with the number of disuade (disconnected) nodes 
negligible compared to $N$. In order to get a non subjective comparison of $\epsilon_c$ between 
different systems, we first transform the time series into a uniform deviate so that the size of the 
embedded attractor is rescaled into the unit interval $[0,1]$. The RN is constructed from the time 
series for $M = 3$ taking a range of $\epsilon$ values from $0.02$ to $0.2$. We then compute the count of 
disuade nodes in the RN as a function of $\epsilon$. The results are shown in  Fig.~\ref{fig2} for 
random and Lorenz attractor time series for $N = 2000$ and $5000$. Note that a giant cluster appears 
for both systems with $< 0.5\%$ disuade nodes corresponding to $\epsilon \sim 0.1$ for $N = 2000$. 
For $N = 5000$, this critical value of $\epsilon$ gets slightly decreased. Hence we use 
$\epsilon_c \equiv 0.1$ as the critical threshold for construction of RN in this paper. 

\begin{figure}
\begin{center}
\includegraphics*[width=16cm]{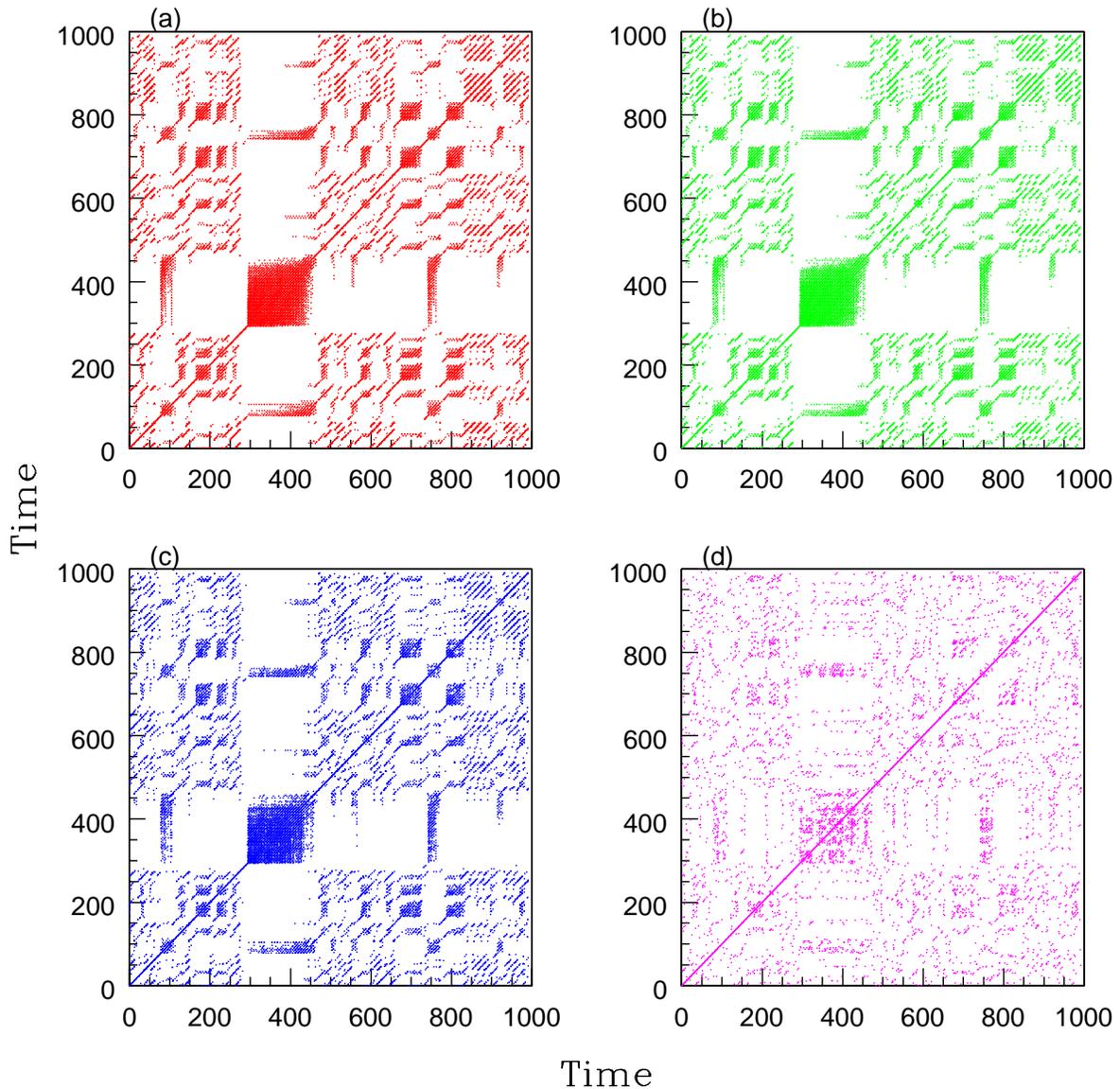}
\end{center}
\caption{The RP constructed from the Lorenz attractor time series by adding four different 
percentages of white noise: $(a) 4\% (SNR = 25)$, $(b) 10\% (SNR = 10)$, $(c) 20\% (SNR = 5)$ 
and $(d) 50\% (SNR = 2)$. Note that the structure of the Lorenz attractor, as reflected in 
the RP, is not completely lost untill the SNR becomes 2.}
\label{fig7}
\end{figure}

\begin{figure}
\begin{center}
\includegraphics*[width=16cm]{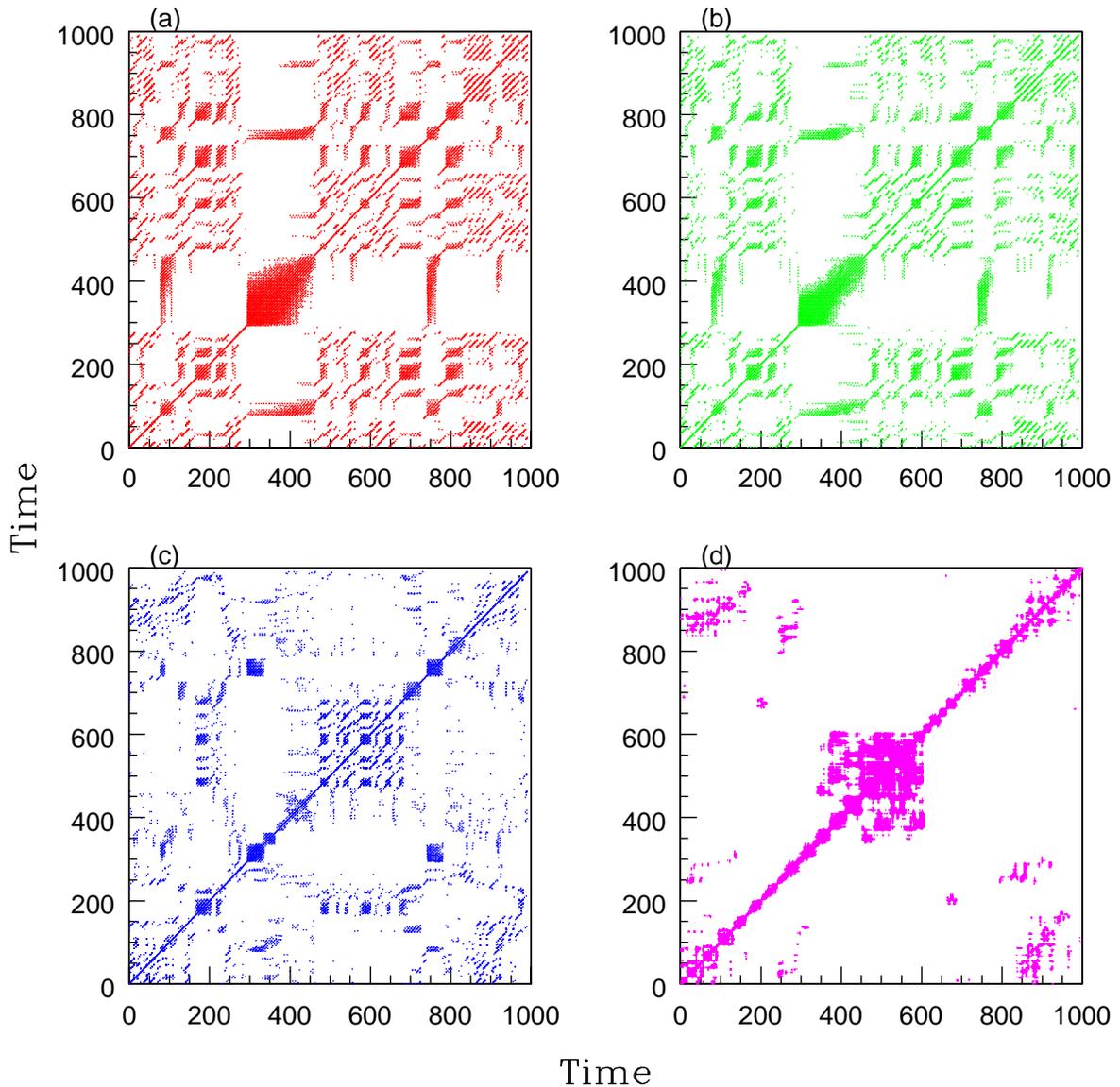}
\end{center}
\caption{Same as the previous figure, but using red noise instead of white noise. As in the 
previous case, only when the noise level reaches $50\%$, the RP loses the characteristic 
features of the Lorenz attractor.}
\label{fig8}
\end{figure}
As an example, we show in Fig.~\ref{fig3} the construction of the RN from the Lorenz attractor time 
series. We use the Gephi software $(https://gephi.org/)$ for all the graphical representations of the 
network in this paper. Note that, by the nature of the network construction, the topology of the 
embedded attractor is preserved in the RN. Each point on the attractor has become a node in the RN 
and the number of nodes connected to a reference node is known by its degree denoted by $k$. 
We show below that the range of $k$ - values can represent the topological changes of the attractor 
due to noise addition and hence we use this as an additional measure, called the $k$ - spectrum, in 
this paper. It is also shown in Fig.~\ref{fig3} for the RN from the Lorenz attractor.

The other important characteristic measures from RN used in this analysis are the degree distribution, 
the characteristic path length (CPL) and the average clustering coefficient (CC).   
The degree distribution indicates how many nodes $n_k$ among the 
total number of nodes $N$ have a given degree $k$. It is usually represented as a probability distribution 
$P(k)$ as a function of $k$, where $P(k) = {{n_k} \over {N}}$. For random networks, $P(k)$ tends to be a 
Poisson distribution for large $N$. The CPL, denoted by $<l>$, is defined through the shortest path $l_s$ connecting 
two nodes $\imath$ and $\jmath$. For unweighted and undirected networks that we consider in this work, 
$l_s$ is defined as the minimum number of nodes to be traversed to reach from $\imath$ to $\jmath$. 
The average value of $l_s$ for all the pair of nodes in the whole network is defined as $<l>$ and the 
maximum value of $l_s$ is taken as the diameter of the network. The CPL can be computed from the equation 
\begin{equation}
<l> = {{1} \over {N(N-1)}} \sum_{i,j}^N l_s
  \label{eq:6}
\end{equation}
The CC of the network is defined through a local clustering index $c_v$. Its value is obtained 
by counting the actual number of edges in a sub graph with respect to node $v$ as reference to 
the maximum possible edges in the sub graph:
\begin{equation}
c_v = {{\sum_{i,j} A_{vi}A_{ij}A_{jv}} \over {k_v(k_v - 1)}}
  \label{eq:7}
\end{equation}  
The average value of $c_v$ is taken as the CC of the whole network:
\begin{equation}
CC = {{1} \over {N}} \sum_{v} c_v
  \label{eq:8}
\end{equation} 
For a detailed discussion of all the network measures, 
see the popular books by Newman \cite {new} and Watts \cite {wat} and some excellent reviews 
on the subject \cite {alb,boc}.

\begin{figure}
\begin{center}
\includegraphics*[width=16cm]{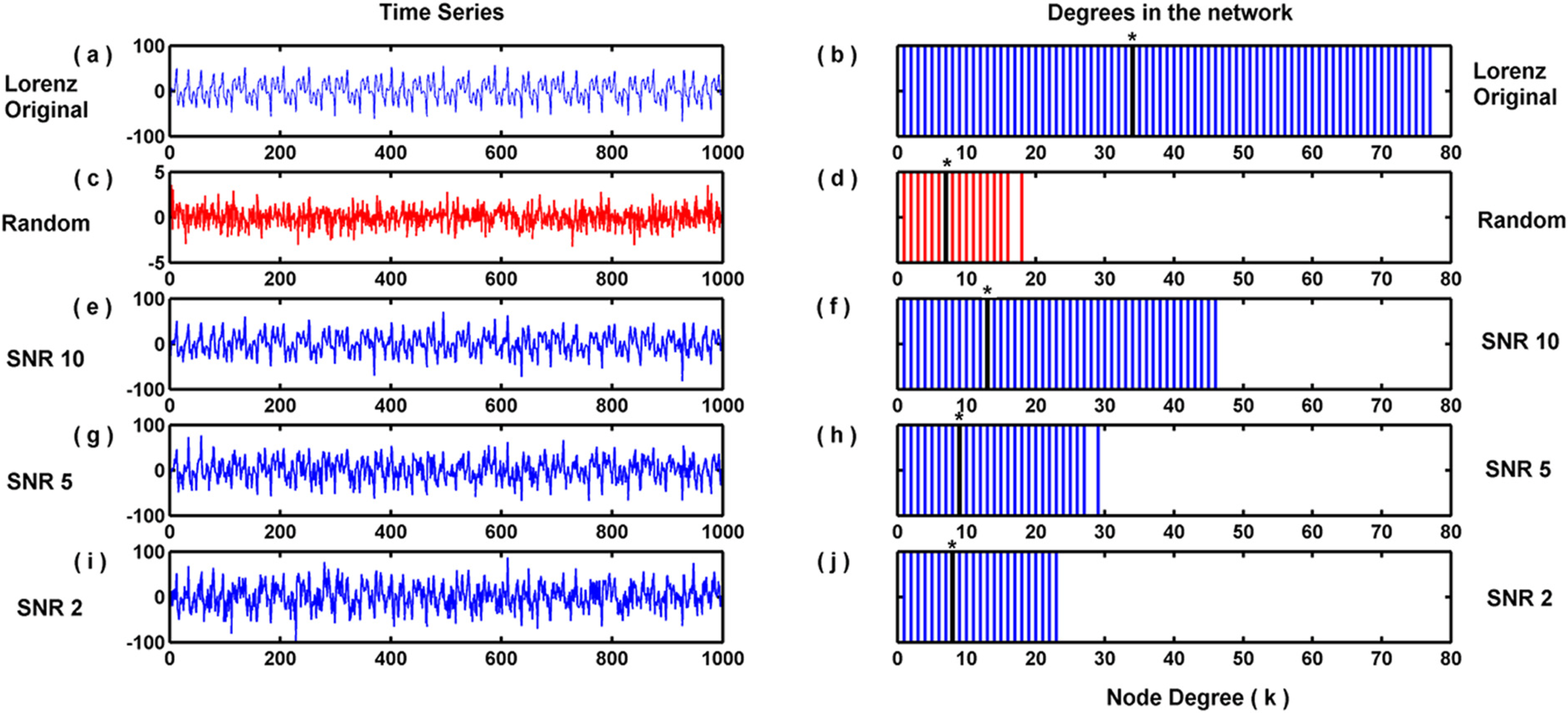}
\end{center}
\caption{The figure shows how the k-spectrum of the RN constructed from a chaotic attractor 
varies by adding different percentages of white noise to the time series from the attractor. 
It is clear that the white noise affects the nodes with high degrees or hubs reducing their 
links so that the network tends to a random network as noise level increases.}
\label{fig9}
\end{figure}

\begin{figure}
\begin{center}
\includegraphics*[width=16cm]{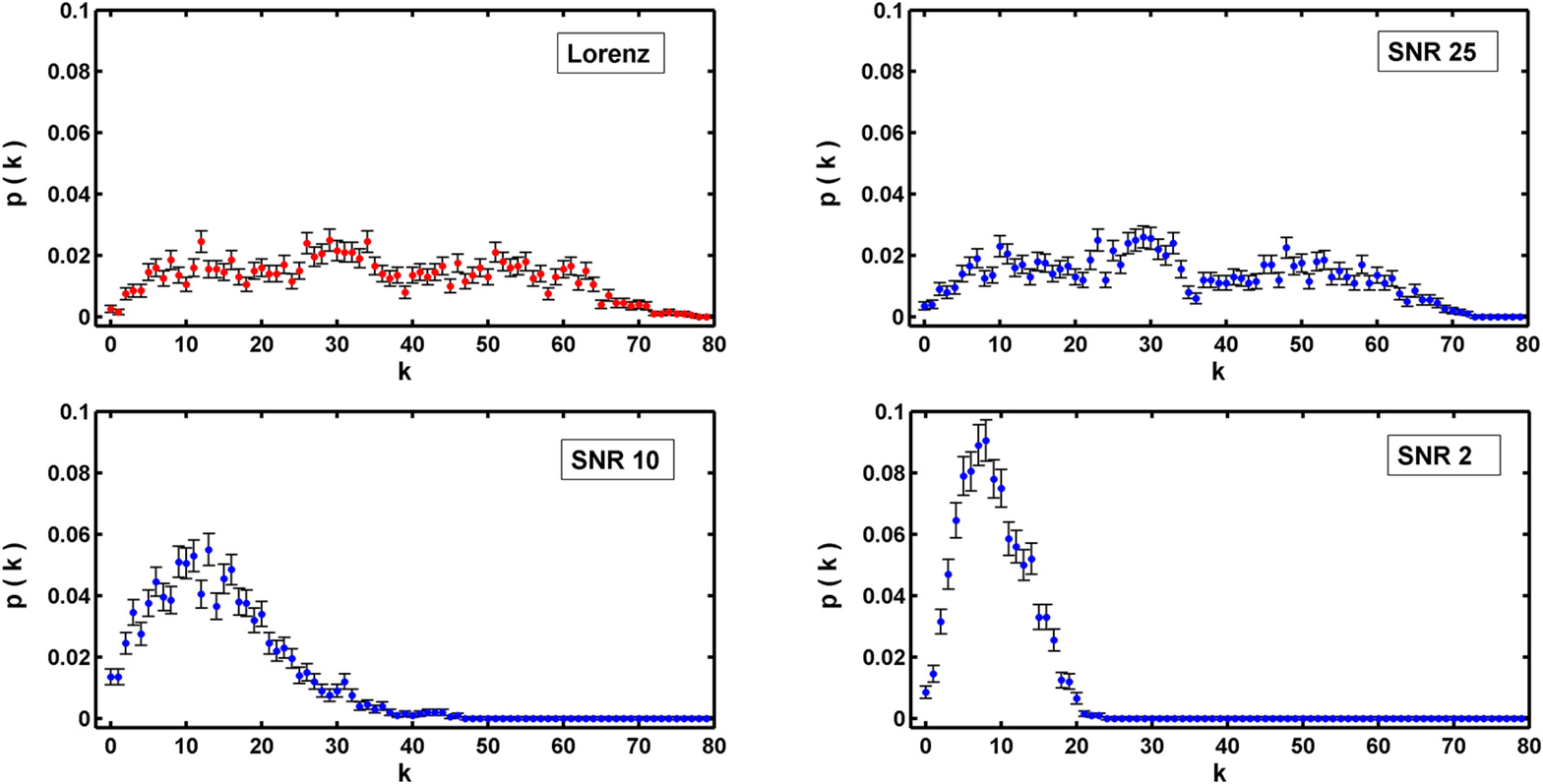}
\end{center}
\caption{The degree distribution for the RNs constructed 
from Lorenz attractor time series by adding different percentages of white noise. The degree 
distribution approaches Poissonian as the noise level increases.}
\label{fig10}
\end{figure}

\section{Analysis of Synthetic Data}
In this section, we study the changes in the topological structure of a chaotic attractor using the 
RN measures by the addition of different percentages of white and colored noise to the attractor 
time series. We use the time series from the standard Lorenz attractor for this analysis. The 
number of nodes in the RN is adjusted to be $N = 2000$ for all computations in this section. 
In  Fig.~\ref{fig4}, we show the Lorenz attractor time series and the constructed RN on the left 
side, where as, the $k$ - spectrum and the degree distribution computed from the RN are shown on the 
right side. The colour gradient is an indication of how the $k$ value changes over the attractor. 
Note that the range of the $k$ - spectrum is from 1 to 78. A \emph{star} over a perticular 
$k$ value in the spectrum indicates that this $k$ value is the most probable or the maximum number of 
nodes in the RN have this degree. It corresponds to the largest peak in the degree distribution 
shown below the $k$ - spectrum. The error bar in the degree distribution is estimated from counting statistics 
resulting from the finiteness in the number of nodes. The statistical error associated 
with any counting of $n(k)$ is $\sqrt {n(k)}$. If $n(k) \rightarrow 0$, one typically 
takes the error to be normalised as 1. Thus, the error associated with $P(k)$ is 
typically ${{\sqrt {n(k)}} \over {N}}$ and becomes $1/N$ as $n(k) \rightarrow 0$. 

Next, we generate an ensemble of noise data sets, both white and colored. The colored noise are 
correlated random fractals whose power vary, in general, as ${{1} \over {f^{\alpha}}}$ with the different 
values of the spectral index $\alpha$ producing different types of colored noise. We use colored 
noise data sets for $\alpha = 1, 1.5$ and $2.0$ for the analysis and results are explicitly shown 
here for one type of colored noise with $\alpha = 2.0$, called the red noise. In Fig.~\ref{fig5}, 
we show the time series, RN, $k$ - spectrum and the degree distribution for the white noise and in 
Fig.~\ref{fig6}, that for red noise. Note that both are completely different with respect to the 
corresponding measures for the Lorenz data. While the RN for white noise does not have a specific 
structure, that for colored noise has a structure similar to that of a chaotic attractor. As 
expected, the range of $k$ - spectrum is much smaller for the white noise with the degree distribution 
tending to a Poissonian, where as, the $k$ - spectrum and distribution of the red noise has a much 
wider range compared even to that of the RN from Lorenz attractor. We have found that the range of 
$k$ values in the RN of colored noise increases as the spectral index $\alpha$ increases from 1 to 2, 
with $\alpha = 1$ very close to that of white noise $(\alpha = 0)$. This difference in the behavior 
between white and colored noise can be attributed to the fact that while white noise tend to fill the 
entire available phase space, the colored noise is a self affine fractal curve which fills only a 
sub space of the total phase space resulting in a highly clustered structure and a large range of 
values in the $k$ - spectrum. 

\begin{figure}
\begin{center}
\includegraphics*[width=16cm]{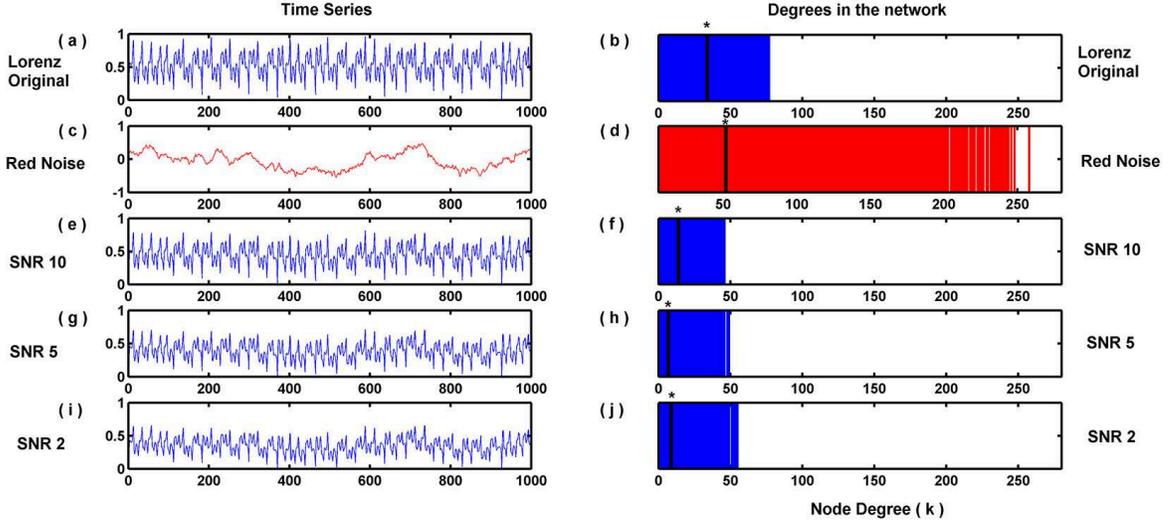}
\end{center}
\caption{The variation of the k-spectrum of the RN constructed from chaotic attractors by 
adding different levels of red noise to the time series from the attractor. Just like the 
white noise, the effect of red noise also is to reduce the high degrees in the network, 
but even with large amount of noise, the k-spectrum does not approach that of pure red noise.}
\label{fig11}
\end{figure}

\begin{figure}
\begin{center}
\includegraphics*[width=16cm]{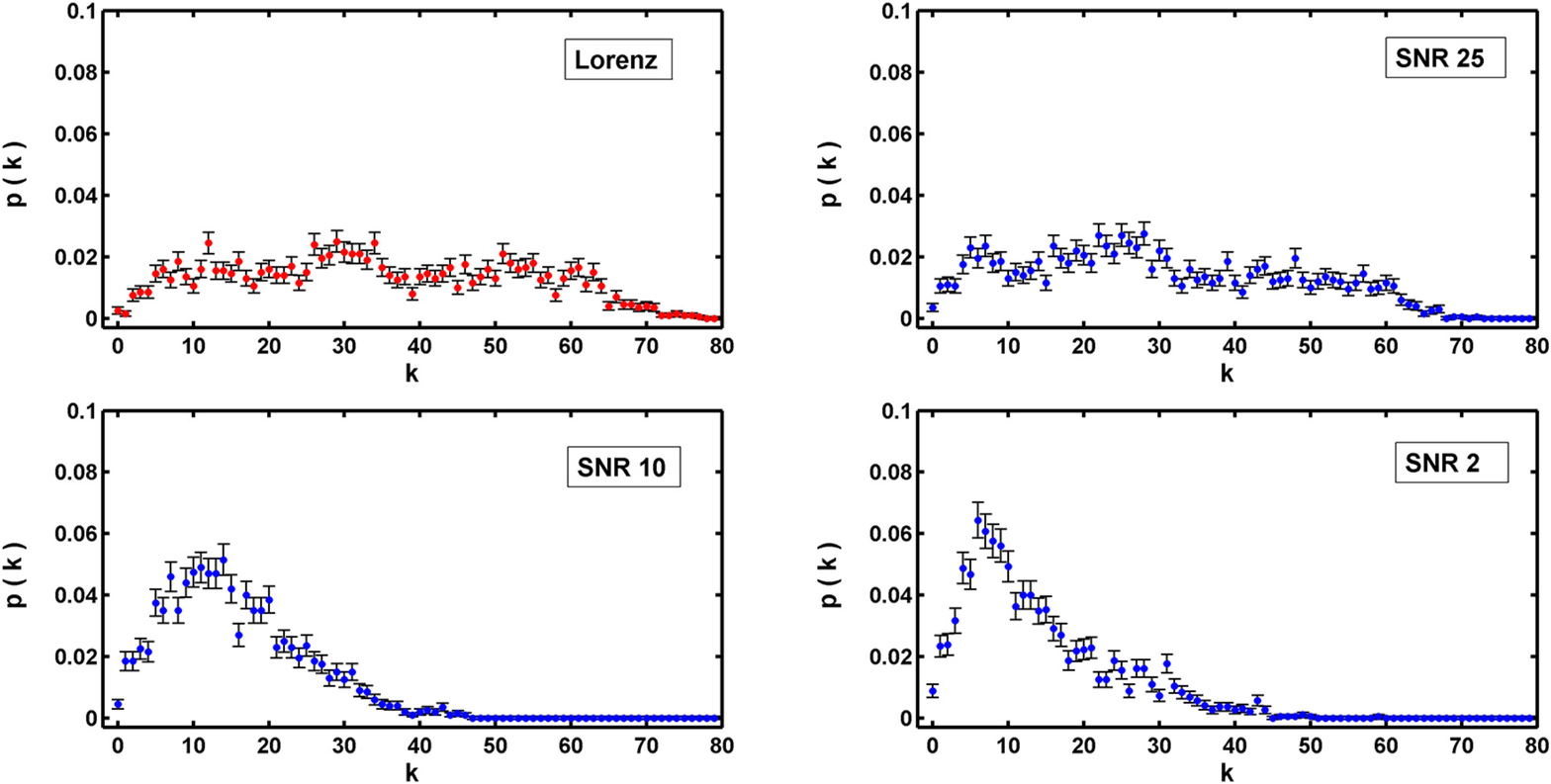}
\end{center}
\caption{Same as Fig.10, but for red noise addition instead of white noise. As the noise level 
increases, the degree distribution deviates from that of the Lorenz attractor.}
\label{fig12}
\end{figure}

To study the effect of noise contamination, we now add different percentages of each type of noise to 
the time series from the Lorenz attractor. Here we show the results for noise addition of $5\%$ (SNR 20), 
$10\%$ (SNR 10), $20\%$ (SNR 5) and $50\%$ (SNR 2). In Fig.~\ref{fig7}, we show the RPs generated from 
time series obtained by adding the above four levels of white noise to the Lorenz data, with the 
noise level increasing from $(a)$ to $(d)$. A visual inspection shows that only when the noise level 
reaches $50\%$, the information in the Lorenz attractor is almost completely lost. For the pure 
Lorenz attractor, the value of $DET$ computed from the RP is $0.988$ while for the pure white noise, 
it is $0.482$. As the percentage of noise increases, the value of $DET$ decreases as $0.961$ for 
$5\%$, $0.902$ for $10\%$, $0.814$ for $20\%$ and $0.604$ for $50\%$. The same results for the 
addition of red noise are shown in Fig.~\ref{fig8}. Again, the value of $DET$ decreases as $0.938$ for 
$5\%$, $0.867$ for $10\%$, $0.768$ for $20\%$ and $0.570$ for $50\%$, with the value for pure red 
noise as $0.455$. Thus it is found that the time series retains much of the information regarding 
the attractor even with moderate noise addition of both white and colored noise.

We now study the effect of noise on the various measures of RN. In Fig.~\ref{fig9}, we show the 
original Lorenz time series and white noise along with time series obtained by adding three different 
percentages of white noise to that of Lorenz on the left panel. The $k$ - spectrum of the 
corresponding RN are shown on the right panel. It is evident that the addition of white noise affects 
the nodes with largest degree or hubs in the RN. As the noise level increases, the range of 
$k$ values keeps on decreasing untill, at sufficiently high noise level, most of the nodes in the 
RN have $k$ value around the average $<k>$, with the degree distribution tending to Poissonian. 
This can be more clearly seen from Fig.~\ref{fig10} where, the $P(k)$ versus $k$ for the four RNs 
in the above case are presented.

\begin{figure}
\begin{center}
\includegraphics*[width=16cm]{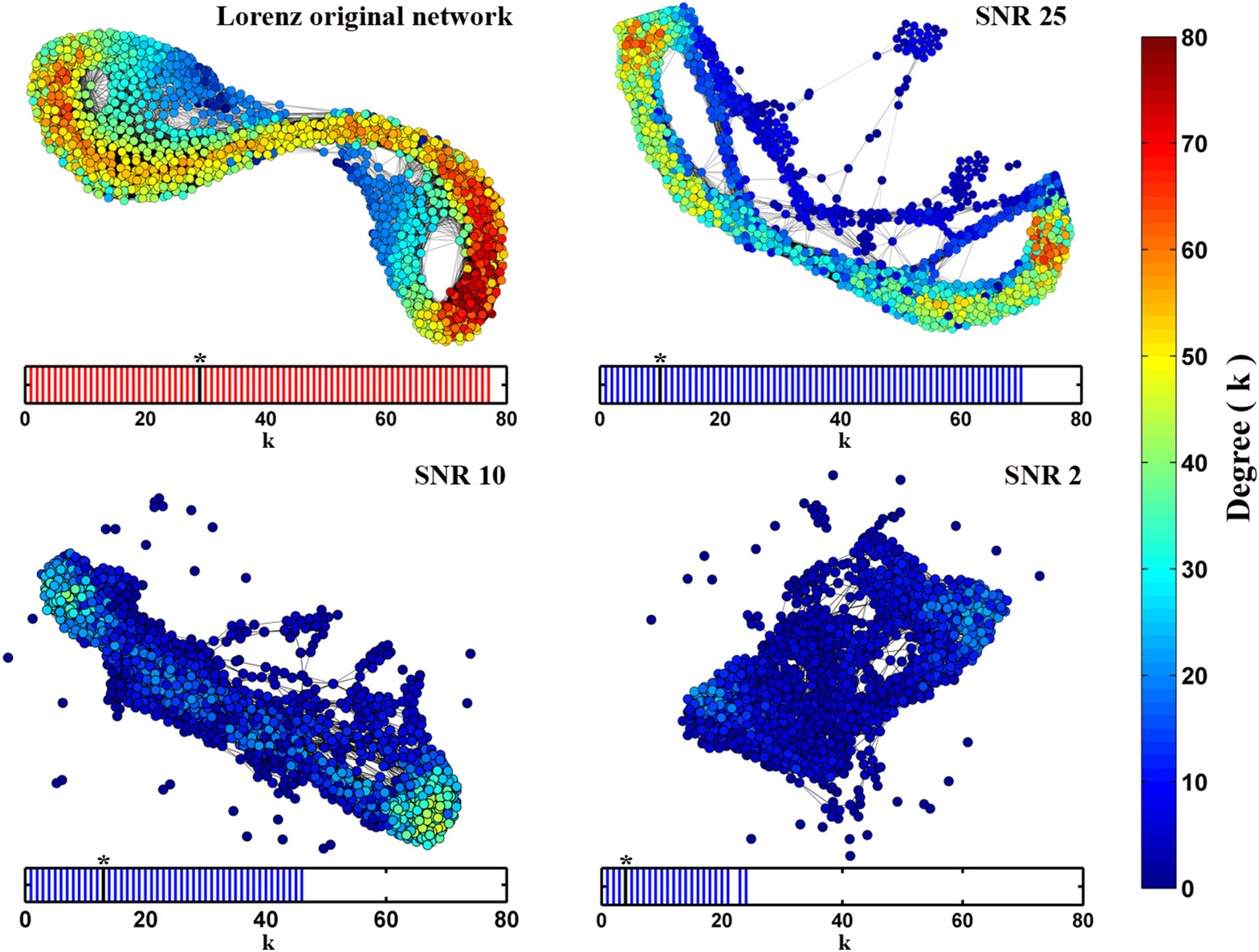}
\end{center}
\caption{The RN and k-spectrum of the Lorenz attractor compared with that of three different 
attractors obtained by adding different percentages of white noise to the Lorenz attractor time 
series. As the noise level increases, the maximum k-value keeps on decreasing. The topology of 
the attractor also approaches that of random.}
\label{fig13}
\end{figure}

\begin{figure}
\begin{center}
\includegraphics*[width=16cm]{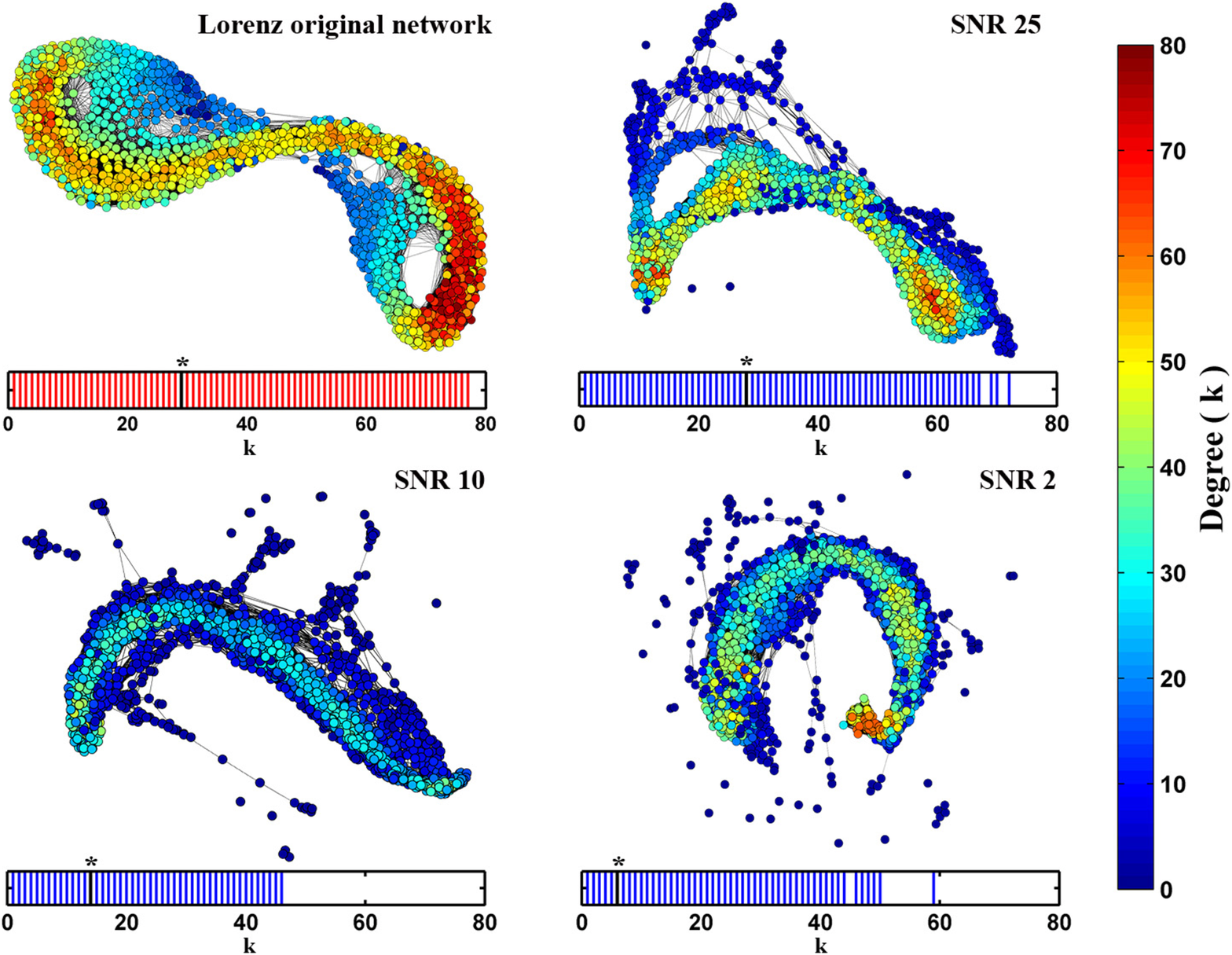}
\end{center}
\caption{Same as the previous figure, but for red noise addition instead of white noise. The 
k-spectrum first decreases and then increases, while the structure changes continuously as 
red noise content increases.}
\label{fig14}
\end{figure}

Since the topology of the RN is reflected from that of the actual attractor, this result implies 
that white noise affects the most clustered or dense regions of the attractor. The noise destroys 
the recurrence of the trajectory points in the phase space and the trajectory points which are 
otherwise, tend to be very close over a time interval, become more widely separated. At sufficiently 
high noise level, the trajectories tend to fill the available phase space completely with 
approximately equal separation between two trajectory ponts.

In Fig.~\ref{fig11} and Fig.~\ref{fig12}, we show the same results as shown in the previous two 
figures, but for contamination with the red noise $(\alpha = 2.0)$. Now the range of the 
$k$ - spectrum is set with that of the red noise as the reference. Here again the range of 
$k$ values gets reduced initially by the addition of colored noise, but tends to increase slowly 
as the noise level is increased. This trend is also reflected in the degree distribution shown in 
Fig.~\ref{fig12}. However, the effect of colored noise is distinctly different from that of 
white noise as is evident from the RP shown in Fig.~\ref{fig8}. The colored noise also destroys 
recurrence of the trajectory and the recurrence occurs only around the main diagonal in the RP. 
This is mainly due to the proximity of the trajectory points in time rather than in space. 
Thus, every node will only be connected to the nodes closer in time and hence have approximately 
the same average degree, except a few nodes having comparatively large degree due to confinement 
and clustering in a lower dimension. This is reflected in the degree distribution shown in 
Fig.~\ref{fig12}. The topological structure of the chaotic attractor in the phase space is also 
changed accordingly.

To show this explicitly, we present the RN for the orignal Lorenz attractor along with that for 
time series added with three different levels of white noise and red noise in Fig.~\ref{fig13}  
and Fig.~\ref{fig14} respectively. The colour grading with respect to the range of $k$ values 
make the figures self explanatory. Note that, in both cases, the structure of the Lorenz 
attractor is not completely lost untill the level of noise contamination reaches $50\%$ in 
agreement with the results obtained from the RPs.

\begin{figure}
\begin{center}
\includegraphics*[width=16cm]{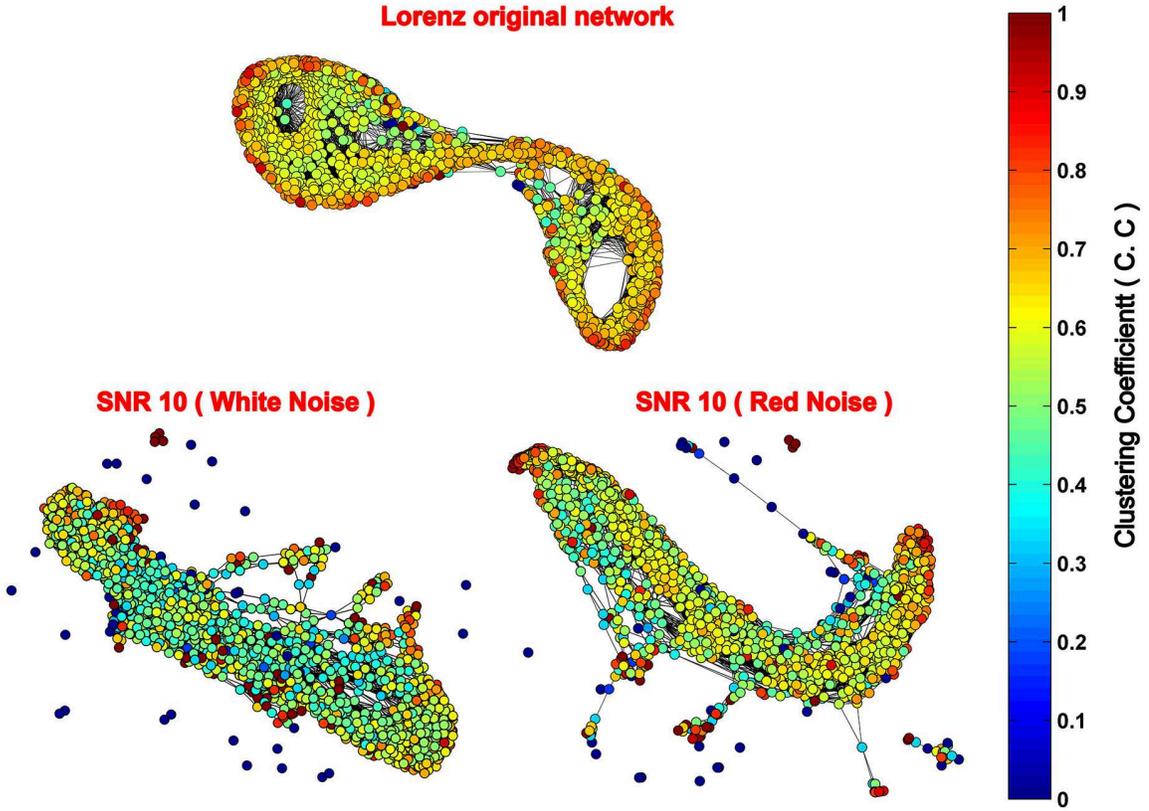}
\end{center}
\caption{Variation of the clustering coefficient of the individual nodes $c_v$ for the RN 
of the Lorenz attractor and that for two attractors obtained by adding $10\%$ white noise 
as well as red noise to the Lorenz attractor time series.}
\label{fig15}
\end{figure}

So far we have been using the $k$ - spectrum and the degree distribution to study the effect of 
noise. It is also important to see how the other network measures vary with the addition of 
white and colored noise. In Fig.~\ref{fig15}, we show how the clustering coefficient of 
individual nodes $c_v$, given by Eq.(7), vary over the RN for three cases, namely, the Lorenz 
attractor time series and that added with $10\%$ of white and colored noise. It is found that 
the $c_v$ of many nodes decrease by the addition of noise, especially the white noise. This 
results in a sharp reduction of the average CC of the RN with the addition of white noise.

\begin{figure}
\begin{center}
\includegraphics*[width=16cm]{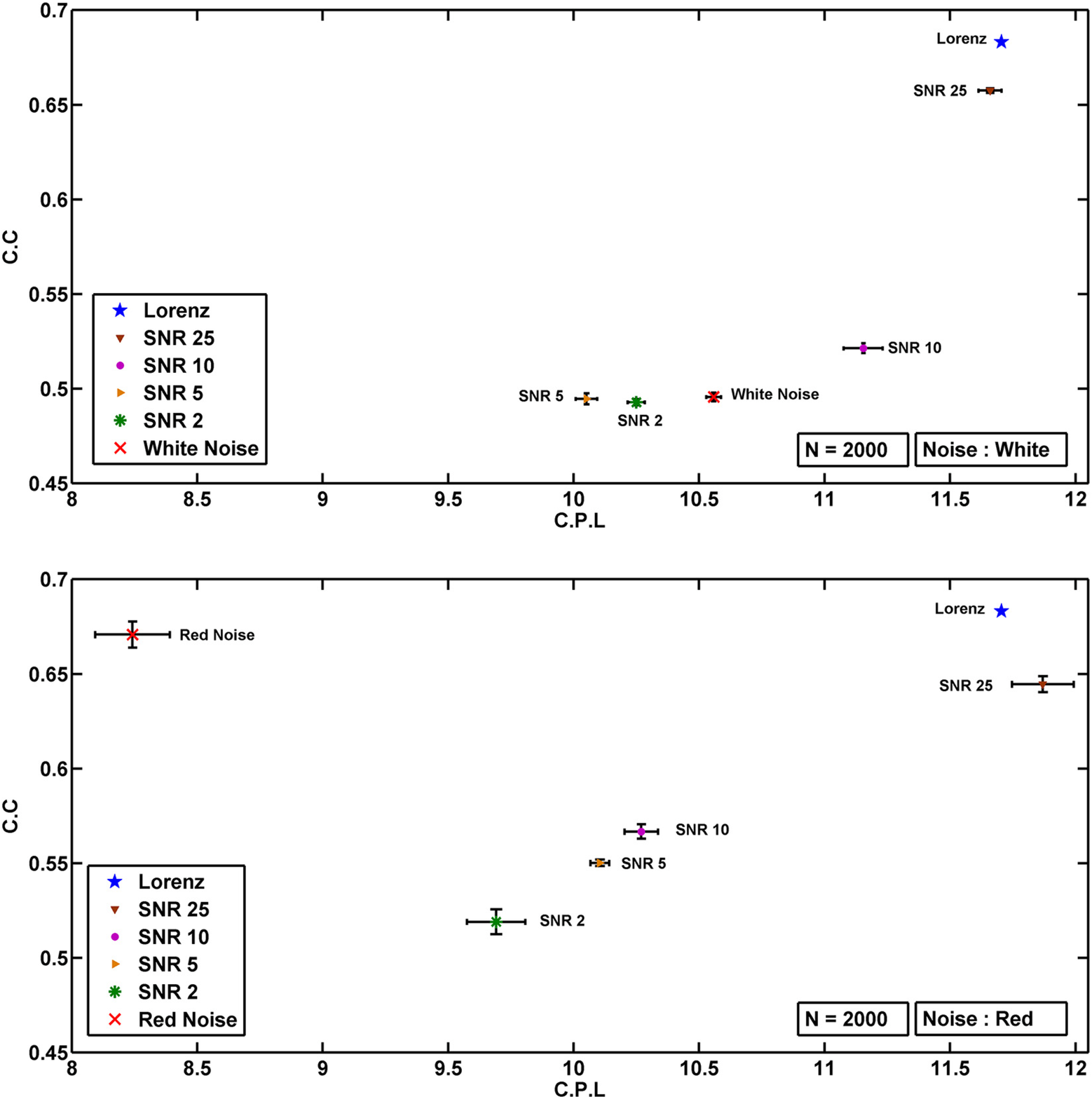}
\end{center}
\caption{The top panel shows the CPL-CC graph for RN from the Lorenz attractor, attractors 
obtained by adding different amounts of white noise to the Lorenz attractor time series and 
RN for random time series.}
\label{fig16}
\end{figure}

Finally, we show how the CPL and the average CC of the RN vary with noise addition by using a 
combined plot. For this, we first generate an ensemble of $20$ time series for the Lorenz attractor 
by changing initial conditions and also $20$ time series each for white and red noise. We then add 
different percentages of white and red noise to the Lorenz data, generating $20$ ensembles of 
time series for each noise level. By constructing the RN from each time series, the CPL and CC are 
computed. The results are plotted on a CPL - CC graph for each type of noise separately. The 
results are shown in Fig.~\ref{fig16} for both white and red noise. The error bar along the 
$X$ and $Y$ directions are the standard deviations in the values for CPL and CC respectively 
from $20$ ensembles in each case.

It is evident that for white noise addition, there is a systematic decrease in the values of both 
CPL and CC as the $\%$ of noise increases. However, the result of addition of red noise is 
different. Though CPL and CC generally decrease with noise, the RN from pure red noise appears to 
behave differently. Its CPL is much less compared to that of all other RNs, but the CC is close to 
that of the RN from Lorenz attractor. The reason for this is not difficult to understand. We have 
found that, for the same number of nodes, the $k$ - spectrum and the range of $k$ values for the 
RN from red noise is much large compared to all other RNs, indicating the presence of some nodes 
with large degree or hubs. This decreases its CPL significantly. Due to the confinement of the 
trajectory in a lower dimension resulting in clustering of trajectory points, the average CC 
remains high.

\begin{figure}
\begin{center}
\includegraphics*[width=16cm]{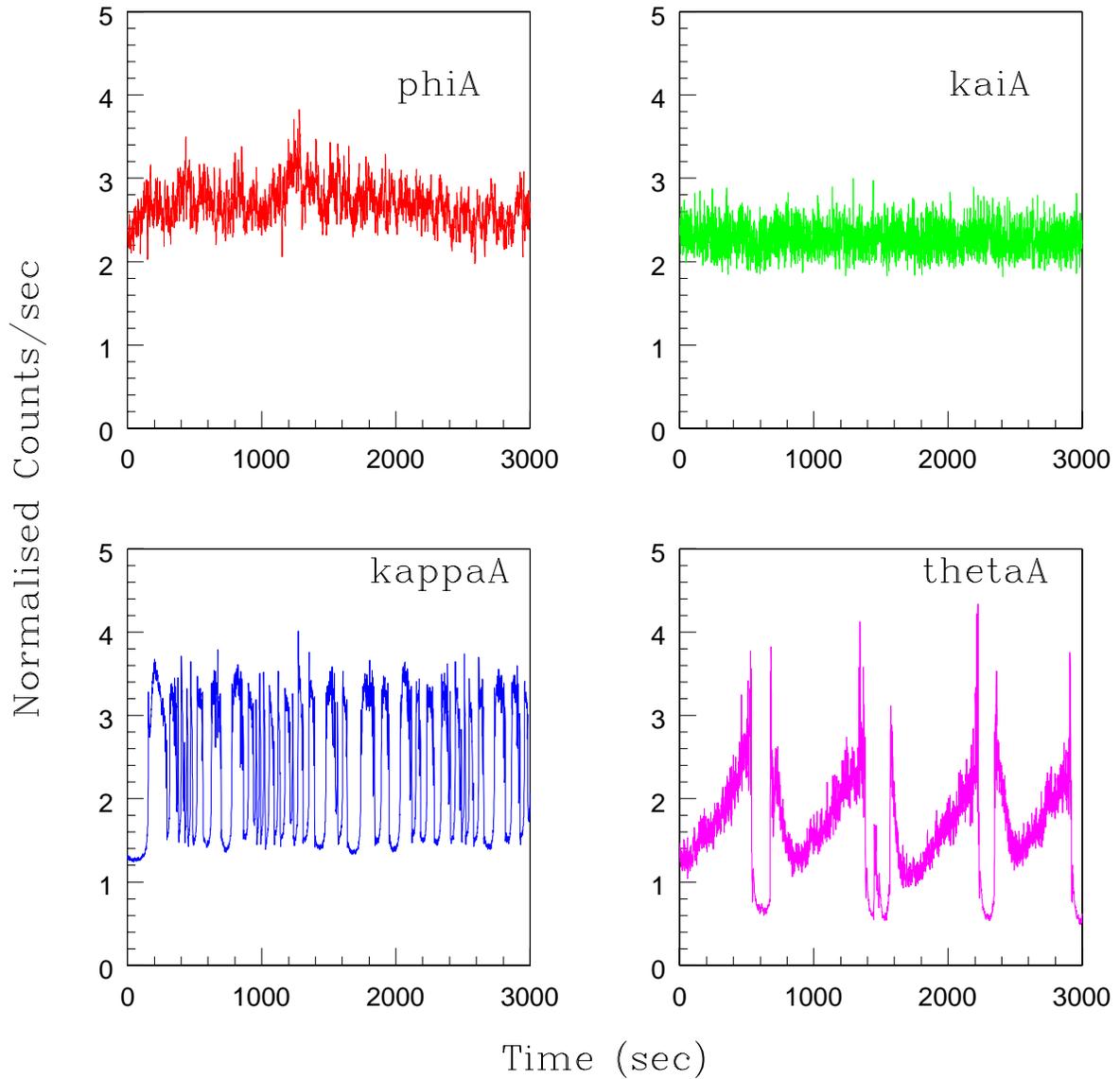}
\end{center}
\caption{The light curves from four representative spectroscopic classes of the black hole 
system GRS 1915+105.}
\label{fig17}
\end{figure}

\begin{figure}
\begin{center}
\includegraphics*[width=16cm]{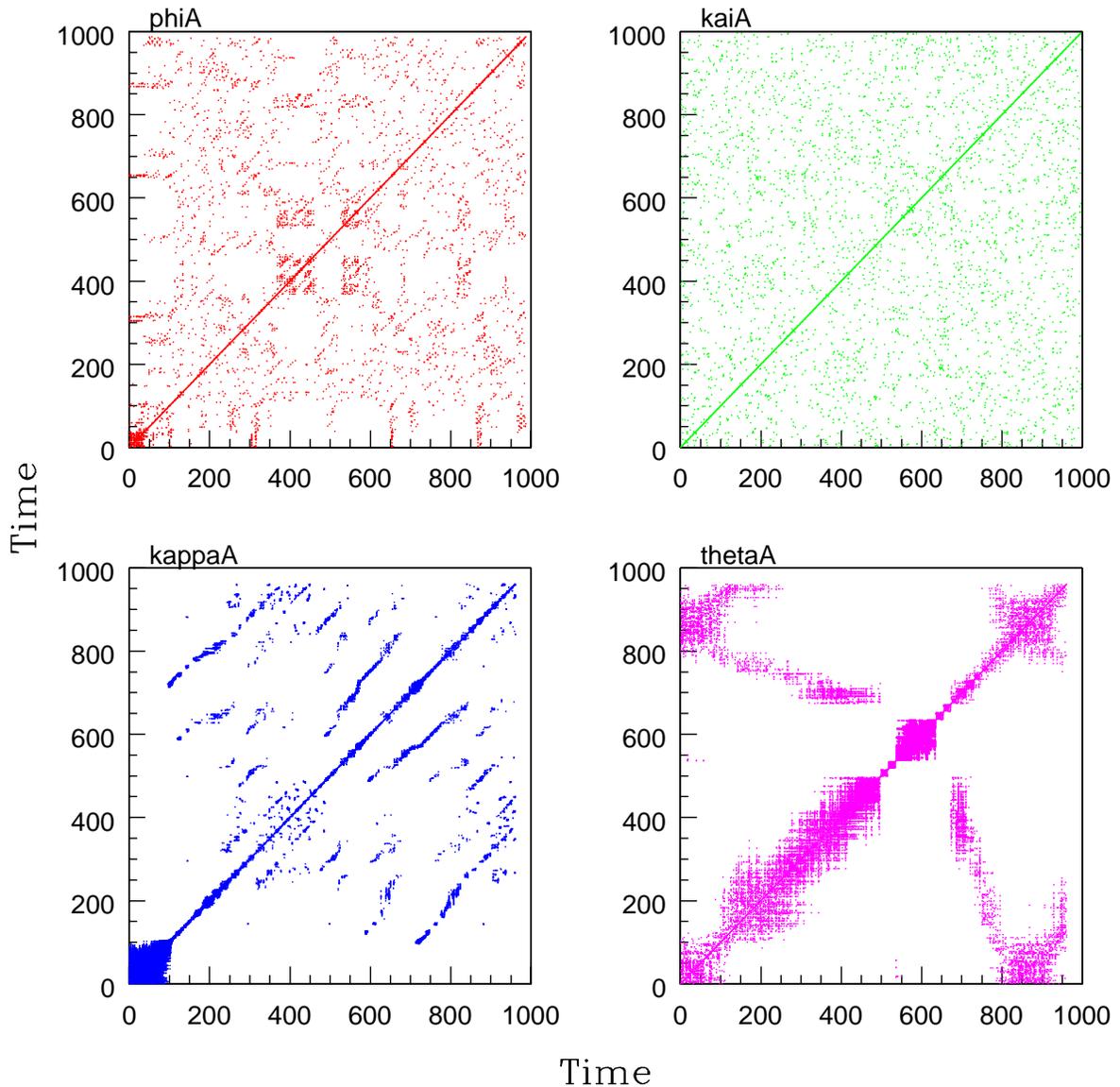}
\end{center}
\caption{The RPs constructed fom the light curves of the black hole system shown in the 
previous figure. The states, except kaiA, appears to have some deterministic nonlinear 
structure mixed with white or colored noise.}
\label{fig18}
\end{figure}

\section{Aplication to Real World Data}
We now check whether the results obtained in the previous section can be used to extract some 
preliminary information regarding the nature observational data from the real world through 
the computation of RN measures. Specifically, we use the X-ray light curves from a dominant 
black hole binary GRS 1915+105 for our analysis. The temporal properties of this black hole 
system have been classified into $12$ different spectroscopic classes by Belloni et al. 
\cite {bel} based on RXTE observation. The system appears to flip from one state to another 
randomly in time. Here we have chosen data sets from $4$ out of these $12$ for our analysis, 
namely, kappaA, kaiA, phiA and thetaA. Each light curve consists of $3200$ continuous data 
points without any gap. We have already applied surrogate analysis to the light curves from 
all the $12$ temporal states and have shown that a few of them show deterministic nonlinear 
behavior \cite {mis}. Recently, by combining the results of computations by various 
quantifiers from conventional nonlinear time series analysis, we were able to group 
some of these light curves together based on their dynamical behavior \cite {kph2}. 
Here we have chosen representative light curves from all these groups for the analysis. 
These light curves are shown in Fig.~\ref{fig17}. 

We first compute the RPs from these light curves and they are shown in Fig.~\ref{fig18}. 
The values of $DET$ obtained from the RPs are $0.723, 0.502, 0.794$ and $0.688$ for 
phiA, kaiA, kappaA and thetaA respectively. A comparison of the figure with  
Fig.~\ref{fig7} and Fig.~\ref{fig8} 
and the values of $DET$ obtained from them tells us that kaiA is close to pure white 
noise while the other three deviate from pure stochastic behavior. PhiA can be considered to be 
contaminated with fair amount of white noise and thetaA is mixed with some form of colored 
noise in high percentage. The bahavior of kappaA suggests that it is a good candidate for 
deterministic nonlinear behavior added with small amounts of colored noise.

\begin{figure}
\begin{center}
\includegraphics*[width=16cm]{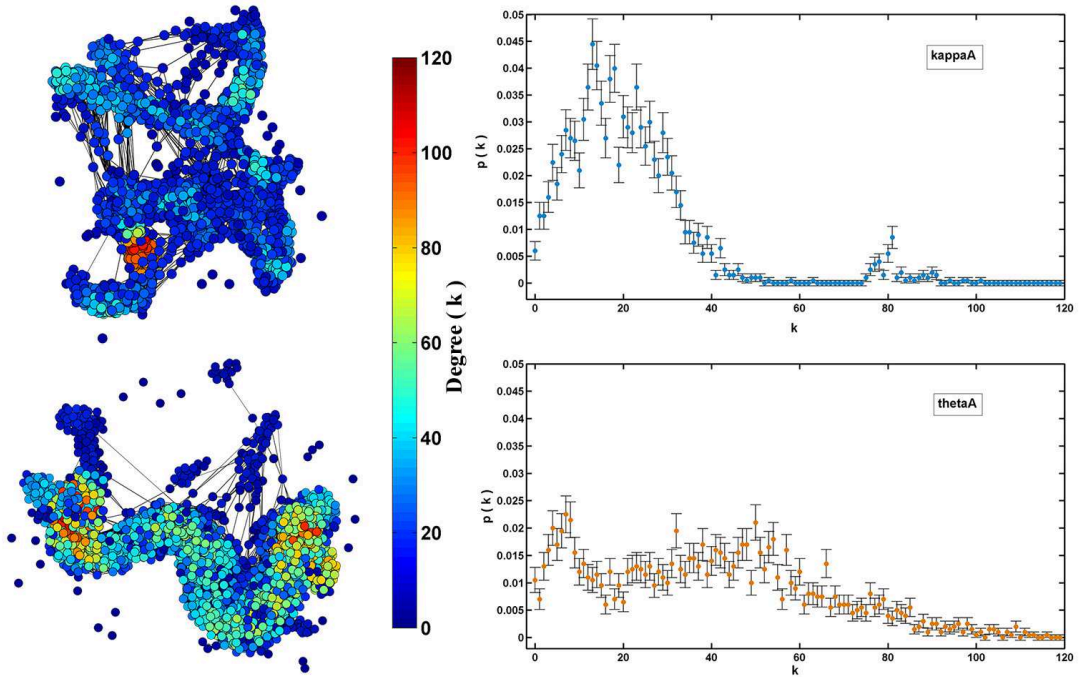}
\end{center}
\caption{The RN and degree distribution for two of the GRS states that appear to have some 
nonlinearity mixed with noise.}
\label{fig19}
\end{figure}

\begin{figure}
\begin{center}
\includegraphics*[width=16cm]{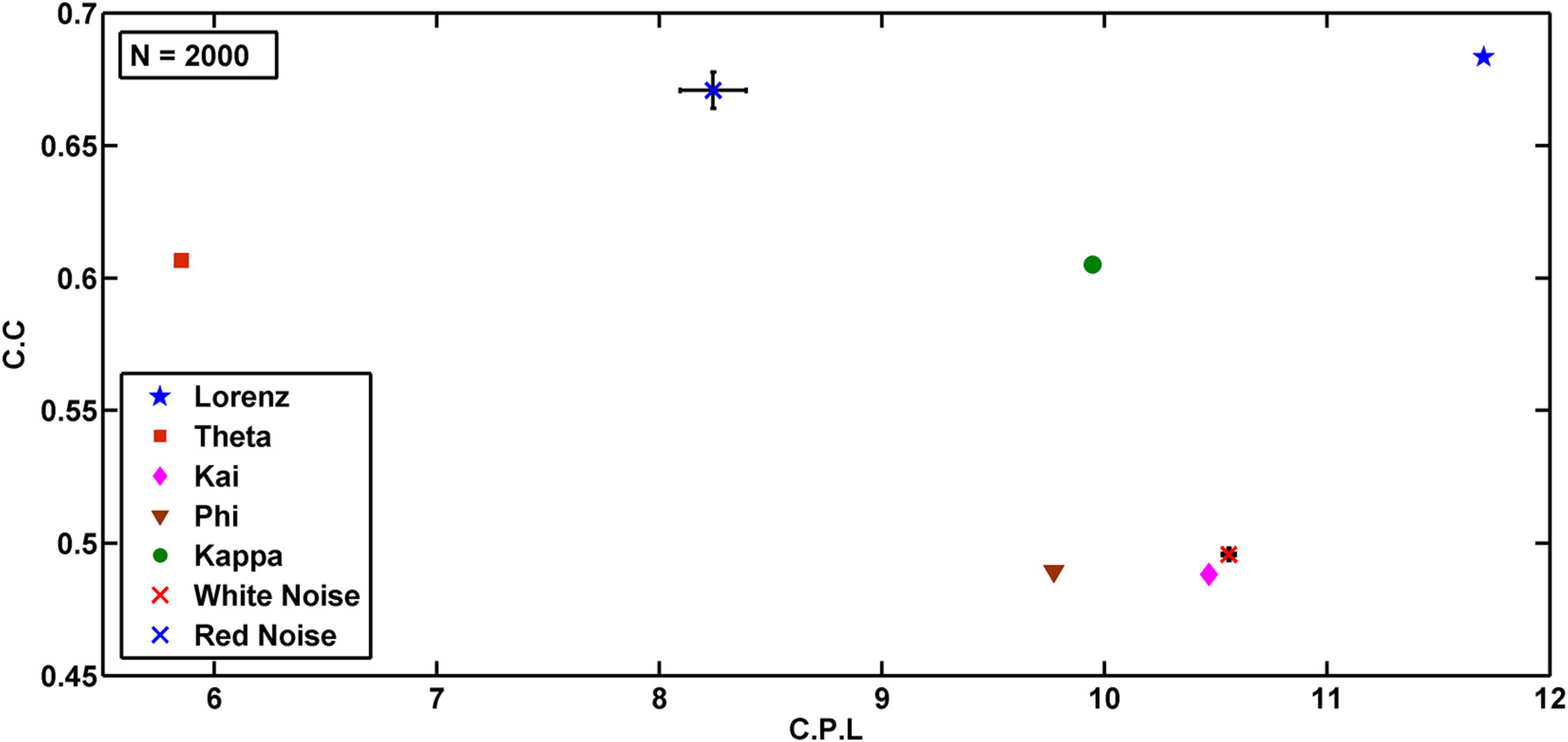}
\end{center}
\caption{The CPL-CC values for the RN from the four GRS states along with that for Lorenz 
attractor, white noise and red noise for comparison.}
\label{fig20}
\end{figure}

To further validate these results, we now compute the RNs and the associated measures from 
these light curves. As expected, RN from kaiA coincides almost exactly with that of white noise 
and RN and distribution of phiA show only small change from that of white noise. However, the RNs 
from kappaA and thetaA are found to be more interesting. These are shown in  Fig.~\ref{fig19} 
along with the corresponding degree distribution. It is evident that both light curves are 
contaminated with colored noise and the $\%$ of noise appears to be much more in thetaA 
compared to kappaA.

Finally, we look at the CPL - CC plot for the light curves as shown in  Fig.~\ref{fig20}. 
For a better comparison, we have added the values for the RNs from the pure Lorenz attractor, 
white noise and red noise. The position of kaiA almost coincides with that of white noise.  
While phiA is slightly away indicating high level of white noise, position of thetaA 
suggests high percentage of colored noise. Only kappaA has least amount of noise 
contamination with network measures comparatively close to that of Lorenz attractor.  

We find that the present analysis has provided us more information regarding the nature of 
noise content in some light curves compared to our earlier analysis using conventional 
measures, such as, correlation dimension and entropy \cite {mis,kph3}. For example, due to 
high white noise content, phiA was identified as white noise in our earlier analysis. The 
present analysis shows that it is not. Similarly, the colored noise content in thetaA was 
not evident in the earlier surrogate analysis. However, our results on kappaA and kaiA 
remain unchanged from previous analysis. Overall, our numerical analysis shows that the 
RP and RN measures are capable of detecting noise contamination in an observational data 
and are especially effective in distinguishing white and colored noise since they change 
the network measures differently.

\section{Conclusion}
There are mainly two purposes for the present study which is numerical in nature. On the one hand, 
our aim is to understand how the addition of white and colored noise to a time series affect the 
topology and structure of the underlying chaotic attractor with the help of RN and the statistical 
measures associated with it. An additional measure, called the $k$ - spectrum, is also introduced by 
us for this purpose. The standard Lorenz attractor time series is used as the prototype for the 
numerical analysis. The method involves construction of the RN from the time series through the 
delay embedded attractor. Different amounts of white and colored noise were added to the Lorenz 
attractor time series and RN constructed. By computing the RN measures, we could show that the 
addition of white and colored noise affect the structure of the attractor differently. However, 
both tend to destroy the characteristic property of recurrence of the trajectory points over the 
attractor. Nevertheless, the characteristic features of the attractor are not completely lost till 
the noise level reaches close to $50\%$ in both cases.

The second purpose of this study is to understand how effective the RN measures are in a noisy 
environment for the analysis of time series from the real world. For this, we use representative 
light curves from different spectroscopic classes of a black hole system GRS 1915+105, where 
both white and colored noise are expected. We compute the measures by constructing the RN from 
these light curves and compare them with the corresponding measures of the RNs obtained from 
the previous analysis with the synthetic data. We show that these measures are effective for a 
first analysis of the data and, to some extent, can distinguish between contamination of 
white and colored noise.

Though the RN and the associated measures have found a large number of applications in various 
fields, to our knowledge, this is the first attempt where these measures are employed to 
study the effect of noise in real world data. A distinct advantage of RN measures is that 
they can be applied to short and non stationary time series, such as, astrophysical 
light curves, physiological signals, etc. Our numerical results obtained here are only 
preliminary. We suggest that a combined surrogate analysis including conventional measures 
as well as the network measures, such as, CC and CPL as discriminating statistic, can provide 
highly accurate results on the nature and amount of noise content in time series from 
real world.

\section*{Acknowledgements}

RJ and KPH acknowledge the financial support from Science and Engineering Research Board (SERB), 
Govt. of India, through a Research Grant No. SR/S2/HEP - 27/2012.

RJ and KPH  also acknowledge the hospitality and the computing facilities in IUCAA, Pune.

\bibliographystyle{elsarticle-num}

\end{document}